\documentclass[showpacs,aps,prl,twocolumn,superscriptaddress,preprintnumbers,nofootinbib]{revtex4-1}
\usepackage{amsmath}
\usepackage{epsfig}
\usepackage{subfigure}
\usepackage{float}
\usepackage{verbatim}
\usepackage{axodraw4j}
\usepackage{pstricks}
\usepackage{color}
\usepackage{slashed}
\usepackage{multirow}
\usepackage{pstricks}
\usepackage{color}
\usepackage{booktabs}
\usepackage{subfigure}
\usepackage{epstopdf}
\usepackage{dsfont}
\usepackage{lipsum}
\usepackage{graphicx}
\usepackage[utf8]{inputenc}

\renewcommand{\section}[1]{{{\textit{ #1.}}---}}

\usepackage{lineno}

\newcommand*{\tmp}[4]{\ensuremath{%
	{#4%
	\ifx\empty#3\empty\ifx\empty#1\empty\else^{#1}\fi\else^{#1(#3)}\fi%
	\ifx\empty#2\empty\else_{#2}\fi}%
}}

\begin{document}
\title{LHC dijet angular distributions as a probe for the dimension-six triple gluon vertex}
\author{Reza Goldouzian}
\author{Michael D. Hildreth}
\affiliation{Department of Physics, 225 Nieuwland Science Hall,University of Notre Dame, Notre Dame, IN 46556, USA} 

\begin{abstract}
In the absence of a concrete discovery of new physics at the LHC, global analyses of the standard model effective field theory (SMEFT) are important to find and describe the impact of new physics beyond the energy reach of the LHC. Among the SMEFT operators that can be constrained via various measurements, the dimension six triple gluon operator involves neither the Higgs boson nor the top quark, yet its variation can have measurable effects on top and Higgs production. Without independent constraints on its impact, the sensitivity of measurements in the top and Higgs sectors to new physics is reduced. 
We show that the dijet angular distribution is a powerful observable for probing the triple gluon operator.
We set the most stringent limit on the triple gluon effective coupling by reinterpreting the results of a search for new phenomena in dijet events using 35.9 fb$^{-1}$ of pp collision data collected at $\sqrt{s}$ = 13 TeV performed by the CMS collaboration. The obtained limit on the strength of the triple gluon operator is far below the sensitivity of that which can be derived from top quark and Higgs measurements and thus this operator can be neglected in global SMEFT analyses. 
\end{abstract}


\maketitle

\section{Introduction}
In the context of standard model effective field theory (SMEFT), a systematic global interpretation of experimental measurements is possible for finding hints of new physics.
The formulation of the SMEFT assumes that new physics lies at a scale much grater than energies accessible at high-energy colliders and can be integrated out from the Lagrangian. In this way, the SM Lagrangian includes higher dimension operators ($O_x$) which are suppressed by powers of the new physics scale ($\Lambda$)~\cite{Grzadkowski:2010es,Buchmuller:1985jz},
\begin{equation}
  {\cal L} = {\cal L}_\text{SM} + \mathcal{L}_\text{eff} = 
    {\cal L}_\text{SM} + \sum_x \frac{C_x}{\Lambda^2} O_x + \dots \ ,
\end{equation}
where $C_x$ stand for the corresponding dimensionless Wilson coefficients.

In the literature, dimension-6 operators  that affect the measurements of  top-quark and Higgs boson properties are studied in detail and are constrained using experimental data collected at the Large Hadron Collider (LHC) \cite{Buckley:2015lku,Ellis:2018gqa,Hartland:2019bjb}.
Among the operators affecting Higgs boson and top quark production processes, the  triple-gluon operator involves neither the Higgs boson nor the top quark. However, this operator can contribute to the total production cross section and in processes involving additional jets and thus its effects have large correlations with other operators involving new physics in the top or Higgs sectors \cite{Buckley:2015lku}.  Not constraining its contribution in global EFT fits can reduce the sensitivity to variations in the operators connected to the top quark or Higgs boson \cite{Hirschi:2018etq,Ghosh:2014wxa}.  This paper presents new constraints on the effects of the triple-gluon operator and proposes an new analysis technique for further refinement of this measurement.

The  triple-gluon operator is the only CP-even dimension-6 genuinely gluonic operator consisting of three factors of the gluon field strength,

\begin{eqnarray}
\label{eq1}
O_{G} =& g_s f_{ABC}G^{A\nu}_\mu G^{B\rho}_\nu G^{C\mu}_\rho 
\end{eqnarray}
Where $G_{\mu\nu}^A=\partial_\mu G_\nu^A-\partial_\nu G^A_\mu+g_sf^{ABC}G^B_\mu G^C_\nu$ and $g_s$ is the QCD coupling.
In principle, the $O_{G}$ operator affects three and four gluon vertices and generates additional vertices with up to six gluons. So one might expect to observe the effects of this operator in inclusive jet production at high energy hadron colliders. However, It has been shown that the helicity structure of the amplitudes for the $gg \rightarrow gg$ and $gq \rightarrow gq$ processes which involve the gluon operator is orthogonal to that of pure quantum chromodynamics (QCD) \cite{Simmons:1989zs,Simmons:1990dh}. Consequently, there is no interference between the QCD and $O_{G}$ operator at $O(1/\Lambda^2)$ and the first non-zero contribution comes at $O(1/\Lambda^4)$. Alternative processes with non-zero interference such as three jet production \cite{Dixon:1993xd} and heavy quark production \cite{Cho1994yu,Sirunyan:2019wka} were suggested in order to constrain the effects of the $O_G$ operator.

Recently in Ref. \cite{Krauss:2016ely}, the authors set a strong constraint on the $C_{G}$ coefficient at 95\% confidence level (CL), $C_{G}/\Lambda^2<0.04$ TeV$^{-2}$, using high-multiplicity jet measurements performed by the CMS collaboration at 13 TeV. They have shown that, although the effects of the interference  terms are negligible in multijet production, terms of order $O(1/\Lambda^4)$ are large enough to be observed in events with high $S_T$ and a large number of jets. In this case, $S_T$ is the scalar sum of jet $p_{T}$s plus any missing transverse energy above 50 GeV. A detailed examination of this analysis \cite{Hirschi:2018etq} concluded that its results are valid and internally consistent even considering 
the contribution of the dimension-8 operators with the same order in an expansion in $1/\Lambda$ as the dimension-6-squared terms and the use of data in the high energy region within the EFT framework. The same considerations and conclusions apply to the analysis strategy presented below.

At the LHC, the production of jets is the process most frequently used to validate the theory of QCD and to search for theories beyond the SM. It is well-known that the dijet angular distributions are an excellent tool to search for new physics \cite{RevModPhys.56.579,PhysRevLett.50.811,PhysRevD.22.184}. Experimentally, the angular variable $\chi_{dijet}$ is defined  as:
\begin{equation}
\chi_{dijet} = exp(|y_1-y_2|)
\end{equation}
where $y_1$ and $y_2$ are the rapidity of the two highest energy jets in the detector frame. 
In SM QCD, the angular distributions  are approximately independent of $\chi$ since all scattering processes are dominated by t-channel gluon exchange. Therefore, new physics contributions that have different production characteristics can be detected on top of the approximately flat angular distributions expected in the SM.    
The dijet angular distributions  are measured at the LHC by the ATLAS and CMS collaborations at 7, 8 and 13 TeV and no significant deviation from the SM prediction is observed \cite{Sirunyan:2018wcm,Aaboud2017yvp}.
Within the EFT framework, the measured dijet angular distributions have traditionally been used to set constrains on the strength of four-fermion operators \cite{Alte:2017pme}.  

In this paper we propose the dijet angular distributions as a powerful observable for probing the triple gluon vertices at the LHC. Furthermore, we use the latest results of the dijet angular distributions measured by the CMS collaboration at 13 TeV to set bounds on the $C_G$ coupling.

\section{Simulation}
We use FeynRules \cite{Alloul:2013bka} to implement the Lagrangian of the $O_G$ operator and write it in Universal FeynRules Output (UFO) files \cite{Degrande:2011ua}. The UFO files are then fed into the Madgraph@NLO Monte Carlo event generator for the event simulation and cross section calculation of the multijet processes \cite{Alwall:2014hca}. Multijet events are generated at leading order (LO) with up to four outgoing partons using the NNPDF3.0 parton distribution function (PDF) set \cite{Ball:2014uwa}. 
The factorization and renormalization scales are set to  the average transverse momentum of the jets.
Generated events are passed to the showering and hadronization performed by PYTHIA8 \cite{Sjostrand:2007gs} with the underlying event tune CUETP8M1 \cite{Khachatryan:2015pea}. The MLM matching scheme is used to remove any double-counting between the matrix element and parton shower calculations \cite{Mangano:2002ea}. Jet reconstruction is performed with the anti-kT algorithm with a distance parameter of 0.4 using the FASTJET package \cite{Cacciari:2011ma}.

Various samples are generated for this study; a SM sample ($C_G=0$), a SM plus $O_G$ sample including the SM-$O_G$ interference, and a pure $O_G$ sample without including SM-$O_G$ interference. The interference effects were evaluated by comparing the prediction of the SM plus $O_G$ (with $O_G$-SM interference) sample to the sum of pure SM and pure $O_G$ samples for distributions of various kinematic observable including $\chi$. As was expected, because of the different helicity structure between the SM and $O_G$ interactions, the  interference effects are found to be negligible compared to the statistical and theoretical uncertainties for $C_G/\Lambda^2=1$ TeV$^{-2}$. The interference effects varies linearly as a function of $C_G$ and become smaller for lower values of $C_G$. Therefore, the interference effects are ignored in this analysis and the pure $O_G$ sample is considered as the signal. The differential jet rate (DJR) distributions are used to check the validity of the merging procedure in the presence of the  $O_G$ operator. The $O_G$ operator does not lead to the soft and colinear divergencies \cite{Englert:2018byk} and DJR distributions are found to be smooth.

\section{Data and SM prediction}
In order to evaluate the power of the dijet angular distributions for probing the $C_G$ coupling, we focus on a recent analysis \cite{Sirunyan:2018wcm} performed by the CMS collaboration at $\sqrt{s}=13$ TeV and with an integrated luminosity of 35.9 $fb^{-1}$. In this analysis, normalized dijet angular distributions, denoted (1/$\sigma_{dijet}$)(d$\sigma_{dijet}$/d$\chi_{dijet}$), are measured over a wide range of di-jet invariant masses and the results are used to probe parameter spaces of various new physics models. We use the public information provided by the CMS collaboration in the HEPDATA database. In particular, we employ the data and SM prediction with the corresponding uncertainties.

Events are required to have at least two reconstructed jets with $p_T>$ 200 GeV and $|\eta|<2.5$. The two highest $p_T$ jets are used to make the dijet system and should have $|y_{boost}|<1.11$, where $y_{boost}= (y_1 + y_2)/2$. Events are further categorized in the following bins of the dijet invariant mass ($M_{jj}$); [2.4,3.0),  [3.0,3.6), [3.6,4.2), [4.2,4.8), [4.8,5.4), [5.4,6.0), and $>$6.0 TeV. The $\chi_{dijet}$ distributions  are normalized to unity in each mass range and then are unfolded to the particle level. 
The contribution of the SM multijet production is predicted at next-to-leading order QCD using NLOJET++4.1.3 \cite{Kluge:2006xs} including electroweak corrections \cite{Dittmaier:2012kx}.

Various experimental and theoretical uncertainties are considered on this measurement \cite{Sirunyan:2018wcm}. In general, the normalized differential cross sections in $\chi_{dijet}$ are relatively insensitive to many systematic effects. The importance of the uncertainty sources varies from the low to high dijet mass bins. In the lowest mass bin, the theoretical uncertainties are dominant, while in the highest dijet mass the dominant source is the statistical uncertainty. The quadratic sum of all systematic uncertainties for all bins of $\chi_{dijet}$ are also provided in HEPDATA and are used in this analysis.

\begin{figure}[!]
  \centering
      \includegraphics[width=0.48\textwidth,height=3cm]{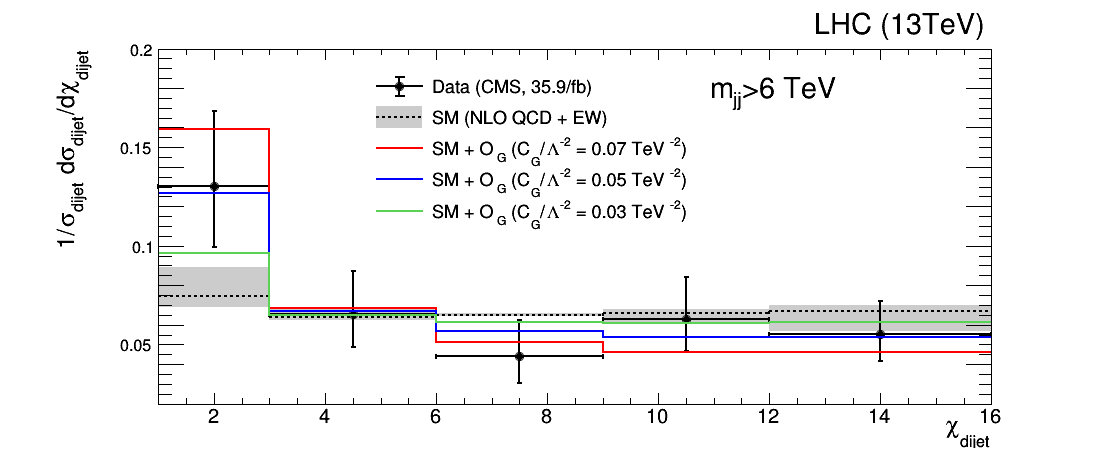}\\
       \includegraphics[width=0.48\textwidth,height=3cm]{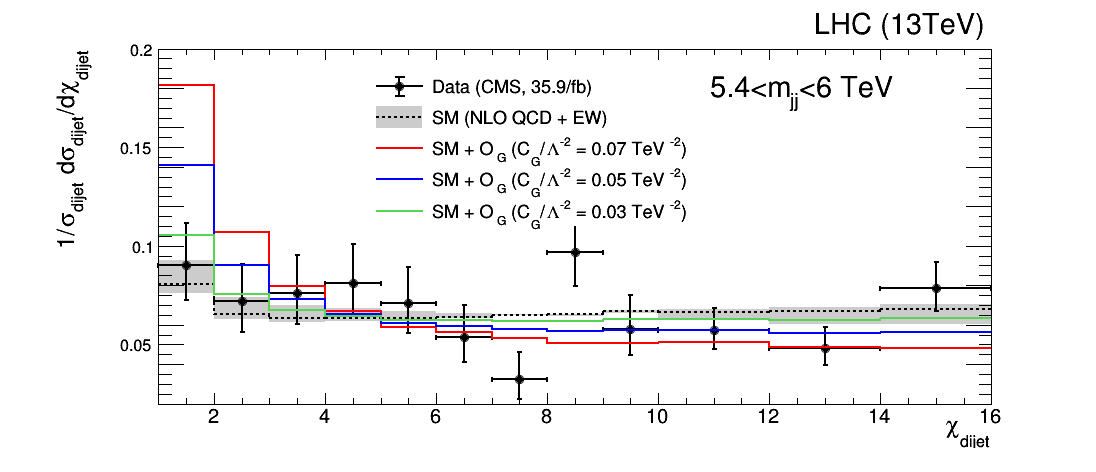}\\
      \includegraphics[width=0.48\textwidth,height=3cm]{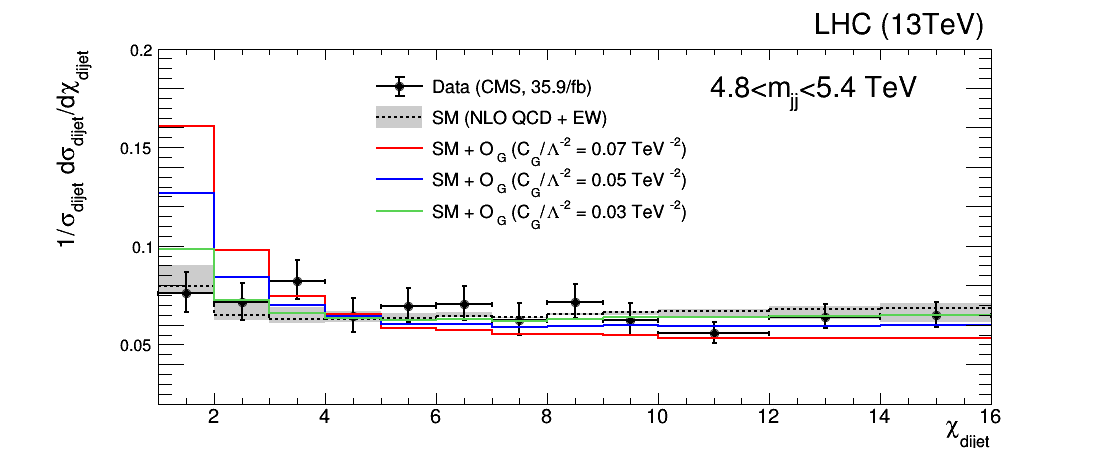}\\
      \includegraphics[width=0.48\textwidth,height=3cm]{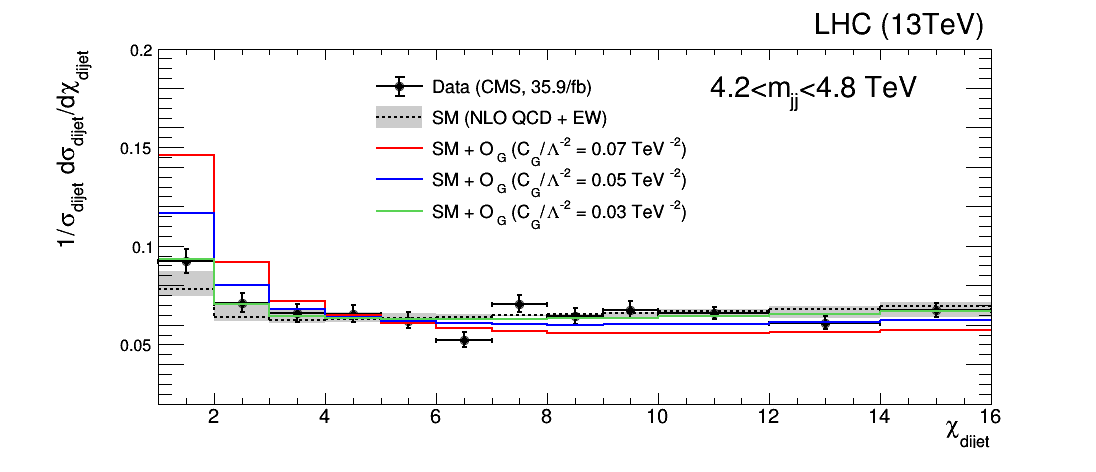}\\
      \includegraphics[width=0.48\textwidth,height=3cm]{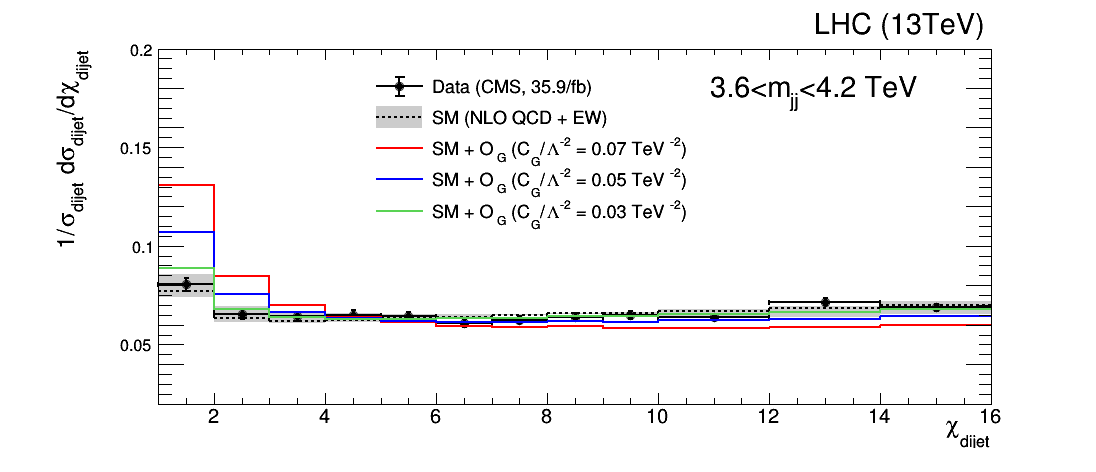}\\
      \includegraphics[width=0.48\textwidth,height=3cm]{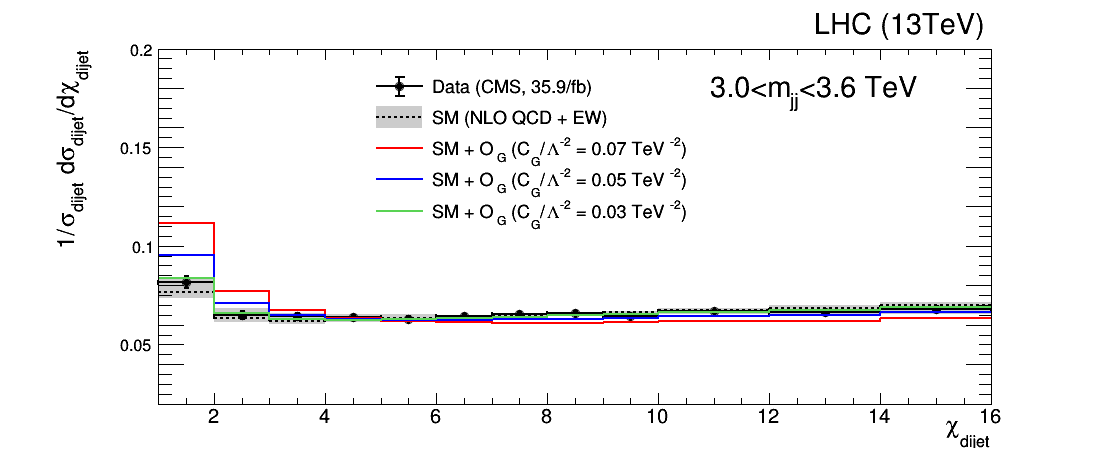}\\
      \includegraphics[width=0.48\textwidth,height=3cm]{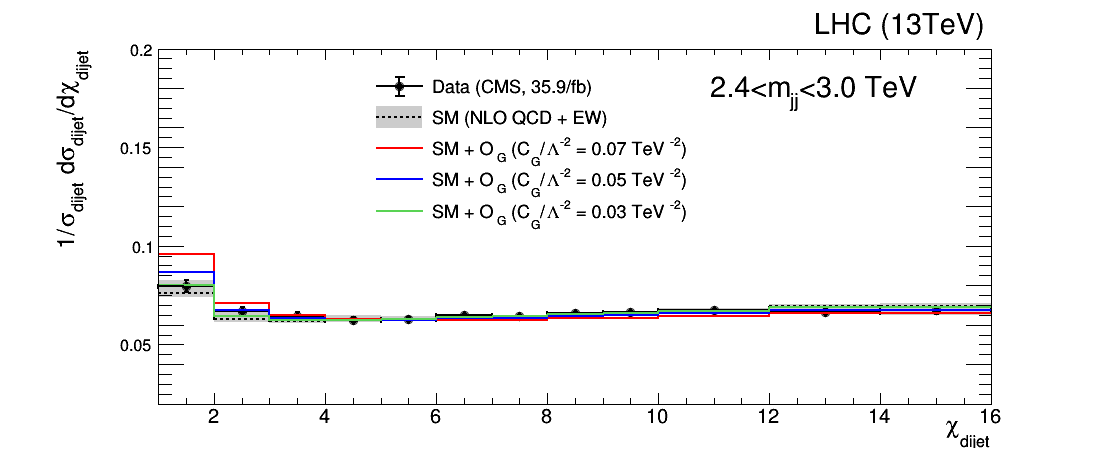}\\
    \caption{Normalized distributions of the dijet angular variable, $\chi_{dijet}$, in different regions of the dijet invariant mass $m_{jj}$. The data (points) and the total theoretical and experimental uncertainties are measured by the CMS collaboration \cite{Sirunyan:2018wcm} and are displayed as shaded bands around the SM prediction. The SM plus $O_G$ expectation is shown for three arbitrary values of the $C_G$ coupling.
    \label{PL-Diff}}
\end{figure}

\section{Results}
In Fig. \ref{PL-Diff}, the normalized $\chi_{dijet}$ distributions of the unfolded data are compared to the SM  predictions and to the predictions of the SM plus $O_G$ for various $C_G$ values in all mass bins.  The data points, SM QCD predictions, and associated uncertainties are imported directly from HEPDATA \cite{Sirunyan:2018wcm}. In order to find the normalized dijet angular distributions in the presence of the $O_G$ operator, we normalize the SM QCD distributions reported by the CMS collaboration  to the cross section predicted by our leading order SM sample described in the previous section. Then, the dijet angular distributions obtained from the pure $O_G$ signal sample are added and the final distributions are normalized to 1. In this way, the template of dijet angular distributions for the SM includes NLO QCD plus EW corrections while the contribution of the signal is predicted at LO.  



It can be seen in Fig. \ref{PL-Diff} that the contribution of the $O_G$ operator peaks at small values of $\chi_{dijet}$, contrary to the SM expectation.  This is due to the fact that none of the $gg\rightarrow gg$, $gg\rightarrow qq$, $gq\rightarrow gq$, and $qq\rightarrow gg$ sub-processes have a t-channel pole \cite{Simmons:1990dh}. Final normalized dijet angular distributions depend on the ratio of the SM cross section to the $O_G$ cross section which varies as a function of the dijet mass. In Table \ref{xs-ratio}, the cross section ratios $\sigma_{O_G}/\sigma_{SM}$ are summarized. The contribution of the $O_G$ operator becomes more important at high masses.
\begin{table}[!tb]
\caption{ The cross section ratio of multijet production with one  $O_G$ vertex to the SM  as a function of $C_G^2/\text{TeV}^{4}$ in the fiducial region for the considered mass bins.}
\label{xs-ratio}
\centering
\resizebox{\columnwidth}{!}{%
\begin{tabular}{|l|l|l|l|l|l|l|l|}
\hline
mass bin (TeV) & {[}2.4,3.0{]} & {[}3.0,3.6{]} & {[}3.6,4.2{]} & {[}4.2,4.8{]} & {[}4.8,5.4{]} & {[}5.4,6.0{]} & $>$6.0 \\ \hline
$\sigma_{O_G}/\sigma_{SM}$  ($C_G^2/\text{TeV}^{4}$) & 13.2          & 24.2          & 38.6          & 53.2          & 67.7          & 89.7          & 112.4  \\ \hline
\end{tabular}}
\end{table}

In order to set constraints on the $C_G$ coupling, we define the $\chi^2$ between data and theory as
\begin{equation}
\label{chi}
\chi^2(C_G) =  \sum_{\substack{ i }} \frac{(x_i^{th}(C_G) - x_i^{data})^2}{\sigma_i^2}
\end{equation}
where $\sigma_i$ is the corresponding uncertainty for the $i$-th bin of the dijet angular distributions. All bins are considered to be uncorrelated for both background predictions and uncertainties. The $\chi^2$  is then minimized and the 95\% CL limit is found for the $C_G$ value at which $\chi^2 - \chi^2_{min}$ = 3.84 \cite{Amsler:2008zzb}. 
The observed (expected) 95\% CL limit on the $C_G$ Wilson coefficient obtained from the combination of all mass bins is 0.031 (0.019) TeV$^{-2}$. 
Based on the expected limits, the most sensitive mass bins are [3.6,4.2), [3.0,3.6) and [4.2,4.8) TeV with the expected limit $C_G/\Lambda^2<0.026$, $0.027$, and $0.028$  TeV$^{-2}$, respectively. The weakest expected limit is found from the highest mass bin ($>$6.0), $C_G/\Lambda^2<0.055$ TeV$^{-2}$ because of the large statistical errors. The obtained limit does not depend strongly on the very high mass bins where the applicability of the EFT might be less valid because of the high energy scales involved. We also found the expected and observed limit combining the  four low mass bins $C_G/\Lambda^2<0.020$ TeV$^{-2}$ and $C_G/\Lambda^2<0.032$ TeV$^{-2}$, respectively.

\section{Conclusions and prospects}
In this paper, we have presented a detailed analysis of the dimension-6 triple gluon operator $O_G$. In principle, anomalous values of this operator could affect Higgs boson or top quark production, making it important to place independent constraints on its value.  It is noted that, to lowest order, there is no interference of the effects of $O_G$ with standard QCD amplitudes in $2\to 2$ processes. This implies that the effects of $O_G$ should be purely additive, making the analsis easy to interpret. We have shown that the dijet angular variable $\chi_{dijet}$ has good sensitivity to anomalous large values of $C_G$, the Wilson coefficient of $O_G$.  Using public data from the CMS experiment, we have set a limit of $C_G/\Lambda^2<0.031$ TeV$^{-2}$ at 95\% confidence level, the most stringent limit to date on $C_G$.  Even given the improved sensitivity of this result, the analysis suffers from constraints related to the use of public data.  The use of relative cross sections for limit-setting cancels some systematic errors but does not give full sensitivity to variations in production cross sections.  In addition, the original simulated signal samples are not available, so some systematic errors could be reduced by careful studies.  We believe, however, that the LHC experiments could produce significantly improved limits on $C_G$ using this technique and the full Run 2 dataset.  In principle, a characterization of $C_G$ should be included in EFT studies so as to independently constrain the value of $C_G$ and its effects on the other dimension-6 operators that can be observed in Higgs boson and top quark events.

\section{Acknowledgment}
We would like to thank B. Clerbaux, H. Bakhshian and A. Jafari for valuable discussions.


\bibliography{main}{}
\bibliographystyle{JHEP}

\end{document}